\documentclass{ws-ijtaf}
\usepackage{color}
\usepackage{float}
\usepackage{graphicx}

\begin{document}

\markboth{Lau Kein Joe}
{Instructions for Typing Manuscripts (Paper's Title)}

\catchline{}{}{}{}{}

\title{GIBBS SAMPLER WITH JUMP DIFFUSION MODEL: APPLICATION IN EUROPEAN CALL OPTION AND ANNUITY}

\author{LAU KEIN JOE 
}

\address{Lee Kong Chian Faculty of Engineering, \\Universiti Tunku Abdul Rahman, \\Jalan Sungai Long, Bandar Sungai Long, 43000 Kajang, Selangor, Malaysia\\
City, State ZIP/Zone,Country\,\\
\email{banana@1utar.my} }

\author{GOH YONG KHENG}

\address{Lee Kong Chian Faculty of Engineering, \\Universiti Tunku Abdul Rahman, \\Jalan Sungai Long, Bandar Sungai Long, 43000 Kajang, Selangor, Malaysia\\
City, State ZIP/Zone,Country\,\\
\email{gohyk@utar.edu.my} }

\author{LAI AN-CHOW}

\address{Lee Kong Chian Faculty of Engineering, \\Universiti Tunku Abdul Rahman, \\Jalan Sungai Long, Bandar Sungai Long, 43000 Kajang, Selangor, Malaysia\\
City, State ZIP/Zone,Country\,\\
\email{laiac@utar.edu.my} }

\maketitle

\begin{history}
\received{(25 September 2017)}
\revised{(Day Month Year)}
\end{history}

\begin{abstract}
In this paper, we are presenting a method for estimation of market parameters modeled by jump diffusion process. The method proposed is based on Gibbs sampler, while the market parameters are the drift, the volatility, the jump intensity and its rate of occurrence. Demonstration on how to use these parameters to estimate the fair price of European call option and annuity will be shown, for the situation where the market is modeled by jump diffusion process with different intensity and occurrence. The results is compared to conventional options to observe the impact of jump effects.
\end{abstract}


\section{Introduction}	
Financial market are well known to be volatile and can be difficult to predict. Despite its characteristics, investors are still trying to learn and forecast the financial market. The most commonly use method in modelling stock price movement is Brownian motion. An example of application of Brownian motion in pricing option is using the Black Scholes model.
	
In early 90s, Black Sholes model is said to be one of the most favorable methods in calculating option prices. The model is calculated with an expected volatility (implied volatility) to project future prices of the financial assets, where many investors used it to calculate the fair price of an option. However, during 1997 financial crisis, most investors are suffering from major losses due to the drastic jumps in prices, including experienced market investors. This had shown that, the Black Sholes model might be useful to a certain extent, however, it could not handles the extreme events where there are large market movements. 

Here we propose a model aimed to extend the usage of Black Sholes model with jumps. Incorporating jumps in Black-Scholes model is not new. However, current available models assume symmetric jump distributions in both upward and downward directions. In our model we will treat the jumps in the two directions separately.

\section{Literature Review}

Geometric Brownian motion is the basis of Black-Scholes model. Mentioned by Samuelson (1952), the stochastic process of the prices of an asset could be described as a geometric Brownian motion (GBM) in a form of a stochastic differential equation (SDE):   

\begin{equation}\label{eqn1}
dS(t) = \mu S(t) dt + \sigma S(t)dW(t),
\end{equation}

\noindent where $\mu$  is the drift rate or the rate of return;
 $\sigma$ is the volatility of the asset;
 $W(t)$ is a Wiener process,and
 $S(t)$ is the spot price of the underlying assets.

	
The work done by Adeosun (2015), explains that as forecasting or anticipating a market takes much more than a Geometric Brownian motion (GBM), where only the volatility and drift of an asset are considered. Model based on GBM solely will not be sufficient enough to accommodate the market prices. There are several times that the asset prices changed significantly greatly than its calculated volatility.

In the past 40 years, various models had been proposed to reflect the discontinuity and jumps in asset's returns including Merton (1996, 2008, and 1976). 
In his model,
Merton added a jump component into the Black-Sholes formula by using the compound Poisson model \citep{Press}
\begin{equation}\label{eqn2}
 S(t) = S(0)e^{(\mu - \frac{1}{2}\sigma^2)+\sigma W(t) } \prod^{N(t)}_{i=1} e^{Y_i},
\end{equation}
where ${N(t)}$ is a Poisson process;
 ${Y_i}$  is a standard normal distributed random variable;
 $\mu$ is the drift coefficient; and
 $\sigma$ is volatility of the underlying GBM.

Assuming the market follows a Geometric Brownian motion, the additional Poisson process describes the arrival of jump events. The jump event has its own drift and volatility terms that differ from these of the underlying GBM. \cite{Kou} modified the above model, with ${Y_i}$ changing to double a exponential distribution. Kou suggested this would enables one to obtain analytical solutions for most path-dependent options, including barrier options and analytical approximations for American options (\cite{Kou}). 
Kou's double exponential distribtion for $Y_i$ is described by the following equation:
\begin{equation} \label{DoubleExpDistribution}
f_Y(y) = p \cdotp \eta_1 e^{-\eta_{1}y}1_{\{y\ge 0\}}  +  q \cdotp \eta_2 e^{\eta_{2}y}1_{\{y<0\}},
\end{equation}
where
\begin{itemize}
\item $\eta_1 >1, \quad  \eta_2 > 0$,
\item $p, q \ge$ 0 , $p+q = 1$, representing the probability of upward and downward jumps;
\item $ \eta_1 >1$  is required so that $ \mathbf{E}(e^y) <\infty $; and 
\item  $ \mathbf{E}( S(t)) <\infty $.

\end{itemize}

Kou had pointed out two properties of double exponential distribution are of importance for the model. The first one is the leptokurtic or ``fat tail'' feature of the jump size where it inherits the return distribution. This properties is make sense as jump of an instrument is not totally random, but surely depends on the characteristic of the instrument itself. For example, an instrument would not have a drastic jump that is several folds of magnitude difference relative to its spot price.

The second crucial properties the double exponential distribution is its martingale property. This unique property allow closed-form solutions (or approximations) for option pricing problems feasible.
Besides, Kou's model is also has the advantage of being  internally self-consistent and  free of arbitrage in an equilibrium setting.
The model is able to capture empirical aspect of the stock markets and the model parameters are straight forward to calibrate \citep{Zeng}.



Kou shows that jump diffusion model can improve the empirical implications of Black-Scholes model, as it remain and retain its analytical tractability. While Wong and Chan(2006) had shown that jump model can be use to attain a better market pricing, than the Black-Scholes model option pricing method using geometric Brownian motion.

Therefore, this research aim to infuse the jump diffusion model into Black-Scholes model, and and check its applicablitity on European call options and annuity. The next section would discuss briefly about how the jump model's parameter is calculated, and the impact of these parameters on option pricing.

%
%
%
\newpage	
\section{Research Methodology}
\subsection{Data Colletion}
In this research, we had choosen Dow Jones industrial (DJI), NASDAQ Composite 100 (NASDAQ 100), FTSE 100, S\&P 500 and NYSE ARCA OIL \& GAS INDEX (OilGas). We retrieve those market data from Yahoo Finance.

The data collected covering period throughout the 1, January 2005 to 1, January 2015. We separate the extracted data in two different period which is between 2005 to 2010, and 2011 to 2015, where the first period covers year 2007 and 2008, which is known as the economical crisis period. The second period is  Hence we could compare the results from Gibbs sampler for both sets of data.

\subsection{Gibbs sampling method and jump diffusion model parameters}

In this research, we will use the Gibbs sampling method to obtain the values of the parameters ($\mu, \sigma,\mu_{J}$ and $\lambda_{\text{jump}}$ ) in the jump diffusion model \citep{Chan,Gibbs}. 
The fundamental idea of the Gibbs sampling method is using the Bayesian inference from Markov Chain Monte Carlo (MCMC) with Metropolis-Hasting model. Metropolis-Hasting model is an iterating process that updates initialize values towards the distribution with an accept-reject method. While the Gibbs sampling technique is a special cases where we accept every updates for the iterations.

The model begin with a probability mass function (pmf), $\pi$ on a countable set of states, $X$ and a real-valued function, $f(X)$.
Here, both $\pi$  and $f(X)$ are assumed to be complicated and computing their values exactly is intractable and sampling exactly is impossible. Hence, we use the model to sample from $\pi$ approximately, or to approximate the expected value $E[f(X)]$ where $X \sim \pi$ and distributed according to $\pi$.

The Gibbs sampling algorithm are as follow:
\begin{arabiclist}\label{GS}
\item 		We introduce the proposal matrix $Q$. $Q$  is a stochastic matrix, where all its element is positive and sum of each row is equal to 1. $(Q_{ab}=Q_{ba}  ,\forall a,b \in X) $
\item 		Initialize $X_0$ in $X$
\item	For iteration where $i=0,1,2,…,n-1$ :    
\begin{alphlist}[(a)]
\item  	Sample $x$ from $Q(x_i  ,x)$ , such that $x_i$ is a fixed, known variable and $x$ is the sample that range over all possible state, (or can be say as $P (x\mid x_i )=Q(x_i,x)$ )
\item 	Sample a $\mu$ from uniform(0,1),
\item	If $\mu <  \frac{(\tilde\pi (x))}{(\tilde\pi (x_i))}$ ,which is the probability of $x$, then $x_{(i+1)}=x$ , else we reject the newly drawn sample and  $x_{i+1}=x_i.$
\end{alphlist}
\item The output will be a sequence of  $\lbrace x_0,x_1,x_2,… ,x_n \rbrace$. 
\end{arabiclist}
The iteration process will be use in sampling parameters. As each different parameters follow their own distribution, by iterating their posterior distribution, the parameters can be sample out.



We implemented a Gibbs sampler for the interested parameters:  the drift, $\mu$ and volatility, $\sigma$ of the underlying asset, the arrival of jump event, $\lambda$ and the intensity of jump, $\mu_{J}$ and jump’s volatility, $\sigma_{\text{jump}}$ for each and every jump. These parameters will show the behavior and movement of the underlying asset, and hence knowing these characteristic will provide an insight for investors when managing risks.

In the context of parameter sampling with Gibbs sampler, the proposal matrix $Q$ would be the posterior distributions of parameters. We will initialize the initial parameter as $X_0$, and will repeat the sampling process until it is converge. The list of output will be the samples of parameter.

The Table 1 reports the values of the parameters from Gibbs Sampling method from a simulated jump diffusion model.


\begin{table}[h]
\tbl{Testing Gibbs sampler algorithm.}
{\begin{tabular}{|c|c|c|c|} \toprule
	Mean           &   Preset &  Mean     &  Standard Deviation   \\ \colrule
	Drift          &   \hphantom{0}0.1\hphantom{00}    &  0.00198  &  0.43901              \\ 
	Volatility     &   \hphantom{0}0.5\hphantom{00}    &  0.47655  &  0.02244              \\ 
	Jump Intensity &   10.0\hphantom{00}     &  9.51044  &  3.77489              \\ 
	Jump Drift     &   \hphantom{0}0.05\hphantom{0}   &  0.05295  &  0.08975              \\ 
	S.D of jump    &   \hphantom{0}0.025  &  0.16545  &  0.03999              \\ \hline
\end{tabular}}
\end{table}

The results shows the limitations in Gibbs sampler converging for the parameter jump drift and standard deviation of jump.
\subsection{Parameters for double exponential models}

Kou had suggested jump diffusion model using a double exponential model in Eqn \ref{DoubleExpDistribution}. The usage of double exponential requires more parameters for Kou's model. Besides the drift and volatility of the assets, 
$\mu$ and $\sigma,$ Kou model requires $\eta_1$ and $\eta_2 $ that replace the $\mu_{J}$ with normal distribution. $\eta_1$ and $\eta_2$ is the mean intensity of upward and downward jump that is needed for double exponential model. The  arrival of jump event, $\lambda_{\text{jump}}$ includes both upward and downward jump event in Kou model.

Throughout the research, we discover that the arrival of upward and downward jump is not dependent to each other. Hence we propose a modified method that splits the parameter of jump arrival into two different parameter. This will ensure the retractability of upward and downward jump arrivals.

\begin{equation}
 S(t) = S(0)e^{(\mu - \frac{1}{2}\sigma^2)+\sigma W(t) } \prod^{N_1(t)}_{i=1} e^{Y_{1,i}} \prod^{N_2(t)}_{i=1}e^{Y_{2,i}},
\end{equation}
where ${N(t)}$ is a Poisson process;
 ${Y_i}$  is a standard normal distributed random variable;
 $\mu$ is the drift coefficient; and
 $\sigma$ is volatility of the underlying GBM.
\begin{equation} \label{myEqn}
\begin{split}
f_{Y_{1}}(y) =  p \cdotp \eta_1 e^{-\eta_{1}y}, \\
f_{Y_{2}}(y) = -q \cdotp \eta_2 e^{-\eta_{2}y}.
\end{split}
\end{equation}

The values of parameters are attain by using different methods as Gibbs sampling method had limitations against noise and drift of assets. First of all, we will take the difference in price for consecutive days. By arranging to ascending order, we can obtain a median value for both positve and negative differences. Value with differences four times larger than the positive median (or four times smaller than negative median) are suspects to be the jump spikes. These will be the jump intensity which will be input into the double exponential part of the modified jump model. The number of times of such spike values are treated as arrival of jump events and will be use in the Poisson distribution of jump event. Lastly, by taking the average log of rate of return of the underlying asset, we would get the drift and volatility of underlying asset.

For an example,  we use the data from Dow Jones in year 2007. We will get an expected upward spike and five downward spike, with intensity of 0.012 and 0.0107 respectively. The drift and volatility of DJI at 2007 is -0.0043 and 0.1454 respectively.

\subsection{Payoff of European call and variable annuity with jumps}


While the number of jump events, is different for each market instrument too. Generally, there is believe where a market crisis will occur once every ten years. However, jump does not means a drop in prices, an upward jump is also possible. How should the investors group and identify the jump aside from daily normal fluctuation, will be a critical element to consider. This part will be further determine in the future research work.

Since the values of the jump parameters are diffferent for different instruments.  We will setting the range of $o < \lambda < 4$, $0 < \mu_{J} < 0.08$ ,while the initial price and strike price equal to \$100. The drift and volatility  of asset, $\mu$ and $\sigma$ values were fixed at 0.08 and 0.4. We will use this to simulate the jump diffusion process and we calculate the expectation of the payoff for the European call option.



The results had shown that ordinary jump model are undereestimating the payoff. As the jump intensity are normalized with mean zero, the expected payoff from jump spike would be zero. This would contradict to our intial idea where the expected jump event occurance is calculated from the real market price. Hence, we suppose that a modification is needed for the jump diffusion model. Instead of a single jump parameter,$\lambda$ we expand to two different jump component accomodating an upward spike and a downward spike.

\subsection{Annuity with jump diffusion model}
The values of annuity varies on different requested requirement by the annuitant. Larger stream of income and insured guarantee in the future would costs higher premium (or a constant annuity payment) and vice versa. Under the Black-Scholes model assumptions, the dynamics of annuity account value are as follow:

\begin{equation}
dA_t=(\mu-c) A_t dt + \sigma A_t dB_t+kdt .
\end{equation}•
Where $A_t$ and $c$ are denoted as the sub account value at time $t$ and the M\&E (mortality and expense) fee payable continuously respectively while  $k$ is the subsequent contributions to the subaccount. 

	Under the Black-Scholes model, the combined geometric Brownian motion (GBM) with jump event (jump model in Eq.(\ref{eqn2})) is given by the following SDE:

\begin{equation}
dS_t=\mu S_t dt+\sigma S_t dB_t+JS_t dN_t
\end{equation}•

With consideration of extreme event, the sub account value will be modified with jump-diffusion model assumption and becomes:
\begin{equation}
\begin{split}
dA_t	& =A_t  ( dS_t)/S_t -cA_t dt+kdt \\
&   =A_t  (μdt+σdB_t+JdN_t )  -cA_t dt+kdt  \\
&   =(μ-c)A_t  dt+σA_t  dB_t+JA_t  dN_t+kdt
\end{split}
\end{equation}

We consider annuitant has guarantee benefits with roll up premium, consisting both guarantee minimum death benefits and guarantee minimum accumulation benefits (GMDB \& GMAB). The guarantee benefits had a pre-agreed guaranteed interest rate $g \geq 0$, which is chosen such that $g<r$ (feature of GMAB). Hence the guarantee benefits is given as following:

\begin{equation}
G_t    = \begin{cases}
		 A_t  ( dS_t)/S_t -cA_t dt+kdt \\
       	 A_t  (\mu dt+\sigma dB_t+JdN_t )  -cA_t dt+kdt  \end{cases}
\end{equation}

This guarantee benefits resembling an Asian put option where the sub account value, $A_t$ becomes the underlying asset. The payoff function, $P(t)$ will be given as following:

\begin{equation}\label{Eqn8}
P(t)=[G(t)-A_t ]_+ = \max⁡\{ G_t-A_t  ,0\},     \qquad \mbox{for     $t\leq T$} 
\end{equation}
We will calculated and simulate the payoff of annuity with jump model in the market of S\&P 500.

%
%
\newpage

\section{Results and Data Analysis}

\subsection{Gibbs Sampler for Market Indexes}
In methodology we had compute and shown that the Gibbs sampler could retrieve the drift, standard deviation along with jump parameters from jump model. Now we further explore on the other market indexes.
	
We picked Dow Jones industrial (DJI), NASDAQ Composite 100 (NASDAQ 100), FTSE 100, S\&P 500 and NYSE ARCA OIL \& GAS INDEX (OilGas) with two different periods. In the Table 2 we show the results from Gibbs sampler using data from year 2005 October to year 2010 December, while Table 3 shows the results for period between October 2010 and December 2015.  

\begin{table}[h]
\tbl{Comparison between extracted parameters of different indexes between year 2005 and 2010.}
{\begin{tabular}{|c|c|c|c|c|c|} \toprule
	Mean	& DJI	& S\&P 500 &	NASDAQ 100 &	FTSE 100 &	OilGas   \\ \colrule
	Drift          &   \hphantom{0}0.10245 &	\hphantom{0}0.11345 &	 \hphantom{0}0.10625 &	\hphantom{0}0.11171 & 	\hphantom{0}0.22821              \\ 
	Volatility     &   \hphantom{0}0.12574	& \hphantom{0}0.14119	 & 	\hphantom{0}0.17082 &	 \hphantom{0}0.14919 &	\hphantom{0}0.22557            \\ 
	Jump Arrival &   19.33768	& 19.24939	& 14.06721	& 16.59051	& 15.91333          \\ 
	Jump Intensity    &   -0.00407	& -0.00519	& -0.00388	& -0.00475	& -0.01131            \\ 
	S.D of jump    &   \hphantom{0}0.04524	& 	\hphantom{0}0.04711	& 	\hphantom{0}0.05277	& 	\hphantom{0}0.04554	& 	\hphantom{0}0.06655             \\ \hline
\end{tabular}}
\end{table}

\begin{table}[h]
\tbl{Comparison between extracted parameters of different indexes between year 2010 and 2015.}
{\begin{tabular}{|c|c|c|c|c|c|} \toprule
	Mean	& DJI	& S\&P 500 &	NASDAQ 100 &	FTSE 100 &	OilGas   \\ \colrule
	Drift          &   \hphantom{0}0.13043 &	\hphantom{0}0.14949 &	 \hphantom{0}0.17544 &	\hphantom{0}0.06069 & 	\hphantom{0}0.11527              \\ 
	Volatility     &   \hphantom{0}0.12365	& \hphantom{0}0.12797	 & 	\hphantom{0}0.14913 &	 \hphantom{0}0.14301 &	\hphantom{0}0.18886            \\ 
	Jump Arrival &   \hphantom{0}5.23150	& \hphantom{0}6.83121	& \hphantom{0}6.07712	& \hphantom{0}6.08239	& \hphantom{0}12.74748          \\ 
	Jump Intensity     &   -0.00592	&-0.00478	& -0.00568	& -0.00446	& -0.00615            \\ 
	S.D of jump    &   \hphantom{0}0.05644	& 	\hphantom{0}0.05294	& 	\hphantom{0}0.05739	& 	\hphantom{0}0.05095	& 	\hphantom{0}0.04405             \\ \hline
\end{tabular}}
\end{table}

The tables shows that the period between 2005 and 2010 contains more arrival of jump relatively to the period between 2010 and 2015. The first period had a minimum arrival of 14 jumps across the indexes. The intensity of jump however had a smaller scale than what we expected. Hence, we believe that the parameters possess clustering effect betweeen jump arrival, jump intensity and its volatility. As we had limited sample of history data, it is impossible to determine the true parameter values.

As we compare the values of second period in Table3 to the first in Table2, the values for jump intensity and its volatility does not differ by much. Whilst the arrival of jump shows an obvious diferrence where most of the indexes had an average arrival of jump of 6, besides NYSE ARCA OIL \& GAS INDEX. \ref{Fig3} below shows the jump arrival of each index for two different period.

\begin{figure}[!htbp]
\centering
\includegraphics[width=0.9\textwidth]{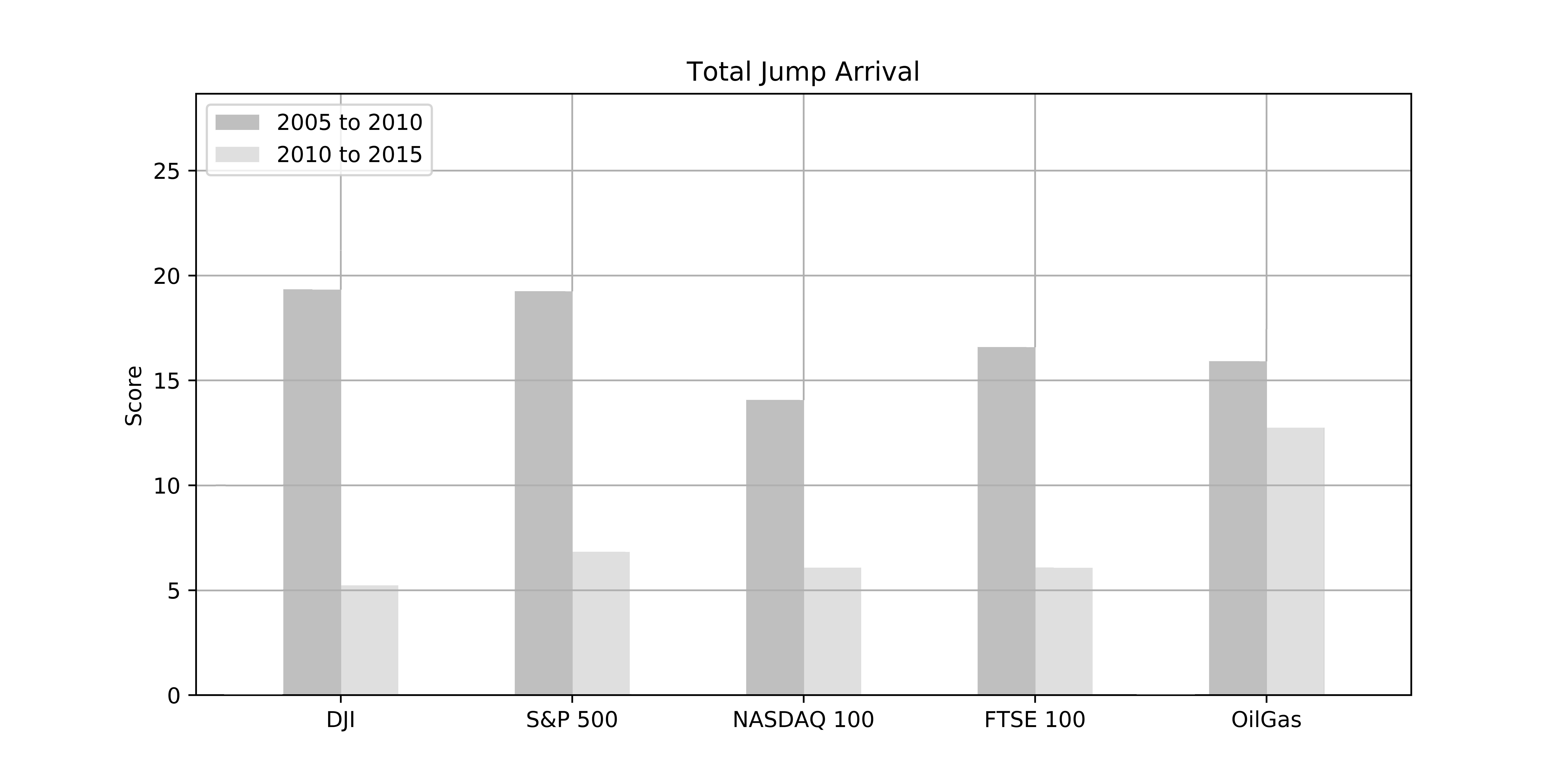}
\caption{Jump arrival of different market index for 2 different period (2005 to 2010 and 2010 to 2015)}\label{Fig3}
\end{figure}

From Figure \ref{Fig3}, we can see that market had a high jump event between 2005 Oct to Dec 2010 compare to the next 5 years period. As we can relate that, we had a world economic recession during year 2007 and 2008. The huge drop in prices caused the jump arrival to shoot up to nearly 20. Whist, the market are more stable after 2010, and hence the jump occurrence are drop to 5 and 6. 

The jumps arrival for NYSE ARCA OIL \& GAS INDEX during the period of 2010 to 2015 are higher compare to the other four indexes. As their average arrival of jump had lower to 6, the oil and gas index remain high at 13 arrivals. Hence, we check on the historical index prices throughout Oct 1, 2005 till Sept 30, 2015 in Figure \ref{Fig4} below.

\begin{figure}[!htbp]
\centering
\includegraphics[width=\textwidth]{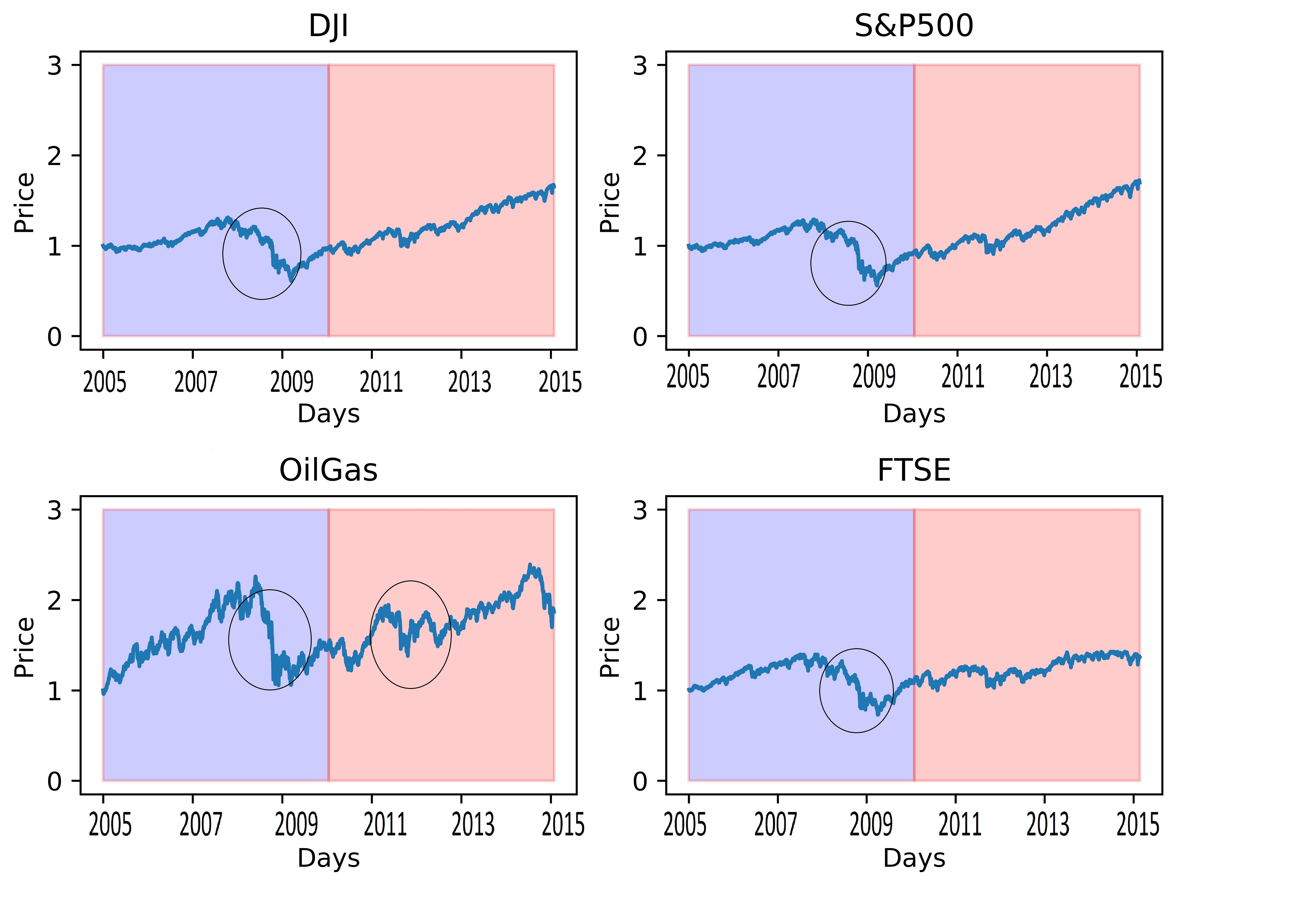}
\caption{Comparison of price behavior for 4 indexes. Data retrieved from Yahoo Finance)}\label{Fig4}
\end{figure}

The Figure \ref{Fig4} above, shows that there are some high spike changes between 2008 to 2009, those region are in the circles. We can see that the NYSE ARCA OIL \& GAS INDEX remains high amount of jump arrival due to the sudden drop in prices on the second period (around 2012), while the other index had started to be slightly more stable. Until year 2015, the oil and petrol prices remain highly volatile from time to time, hence the arrival of jump event remain at a higher level.

This shows that the Gibbs sampler could provide good expectation on the number of jump arrival occured over the period, however it might suffer from clustering effects between the parameters.

In the next section, we will compare jump diffusion model with European call options and annuity, and check its impact on the pricing of both instruments. This is to check whether the jump model could hold its characteristic among two different types of instruments.

\subsection{The Comparison between Jump Diffusion Models and Geometric Brownian Motion Model on Payoff of European Call Option.}
By Blacks-Scholes model, the options pricing could be calculated, to determine the potential risk and return of it. As jump event included in jump model, hence the risk and payoff shall be differ from ordinary Blacks-Scholes. 

The purpose here is to determine how much and how big the impact of jump spike would affect the payoff of the call option. Here,we had built in a jump diffusion model and Geometric Brownian motion model function, where we can simulate its sample. We set fix for the initial price, $S_0$ of 100, with a strike price, $K= 100$ too.  The drift $\mu$ and volatility $\sigma$ be a constant. 

The payoff of the European call option using geometric Brownian motion is calculated using Black-Scholes model. We will exercise the option when the expected price at time $t$, $S_t$ is higher than strike price, $K$. We will not exercise when it is less than that. Hence the payoff at time,$t$ is equivalent to $\max⁡(S_t-K,0)$.

Whilst, the payoff of the European call option using jump diffusion model will be calculated using Black-Scholes model too, similarly to above. However, the jump diffusion model has different ranges for its parameters. For its arrival of extreme events, $\lambda$ or jump arrival in Figure \ref{Fig5} below, it wil be set from zero (no jump event, or normal geometric Brownian motion)
until four, which indicating four spikes per year. While the another parameter is the intensity for each jump is ranging from 0 to 0.8.  The results are shown in Figure \ref{Fig5}.

\begin{figure}[ht]
\centering
\includegraphics[width=0.8\textwidth]{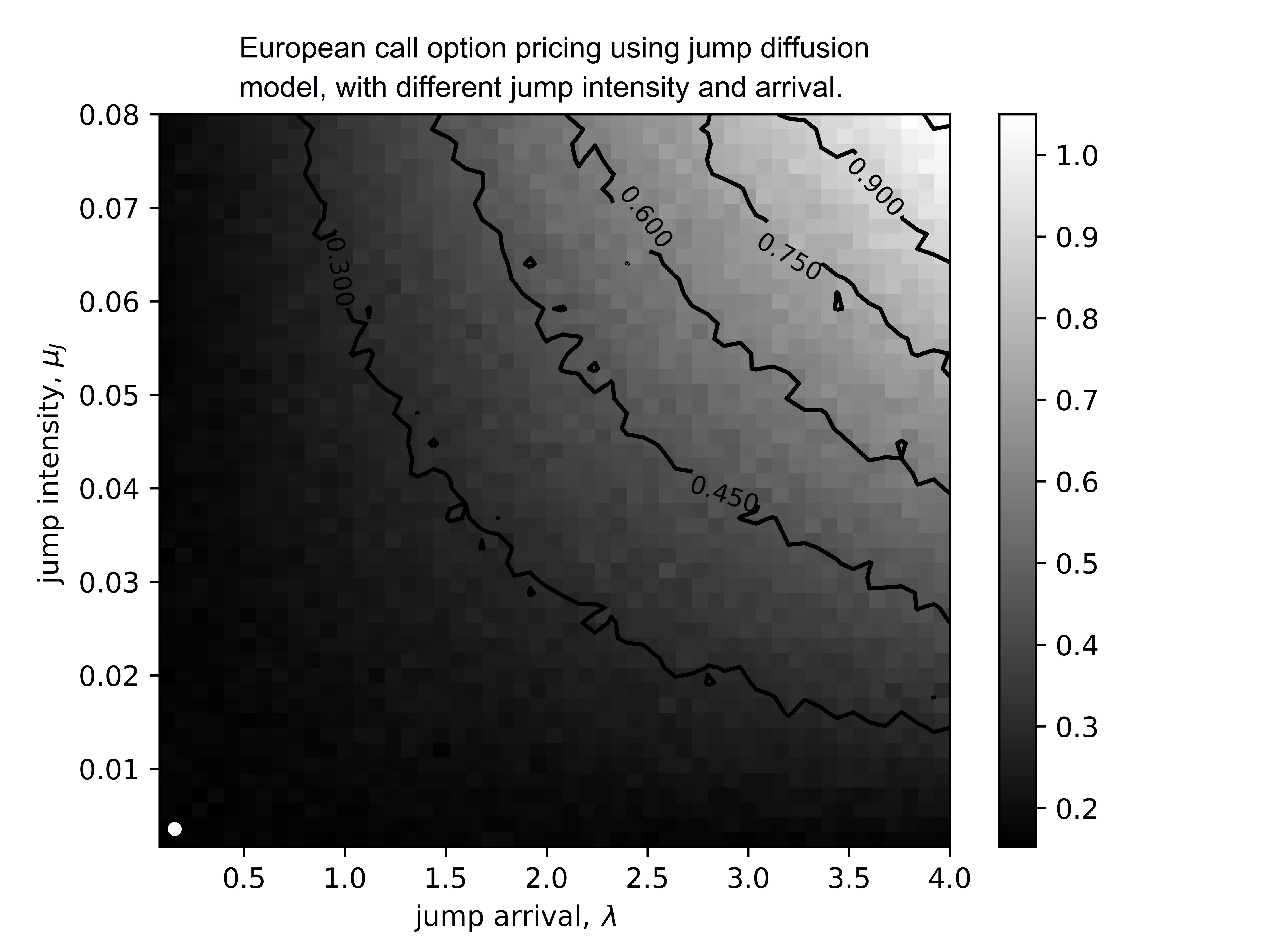}
\caption{The expected payoff for different jump intensity and arrival}\label{Fig5}
\end{figure}

The white dot in Figure \ref{Fig5} is the payoff of European call option with geometric Brownian motion. Along the vertical axis when jump had zero arrival, we can observe that the call option's payoff fluctuates around zero.

When the jump event occurs, the payoff surpass the expected payoff of the Brownian motion at white dot and escalate proportional to the increment of intensity.. From the result, as the intensity of jump increase, the payoff increase more. The highest peak occurs, when the jump intensity and arrival are at their highest level. 

This results only consider if there is positive jump spike, there would be negative jump spike where will induce larger negative payoff for an assets. This shows that with Black-Scholes model, geometric Brownian motion could not expect the larger risk behind each jump event that will occur. This will causes wrong pricing for the call option in the market.

We will take S\&P 500 index and Dow Jones index and compare with the European call option. The figure will show Down Jones index on the right and S\&P 500 index on the left.

\begin{figure}[ht]
\centering
\includegraphics[width=1\textwidth]{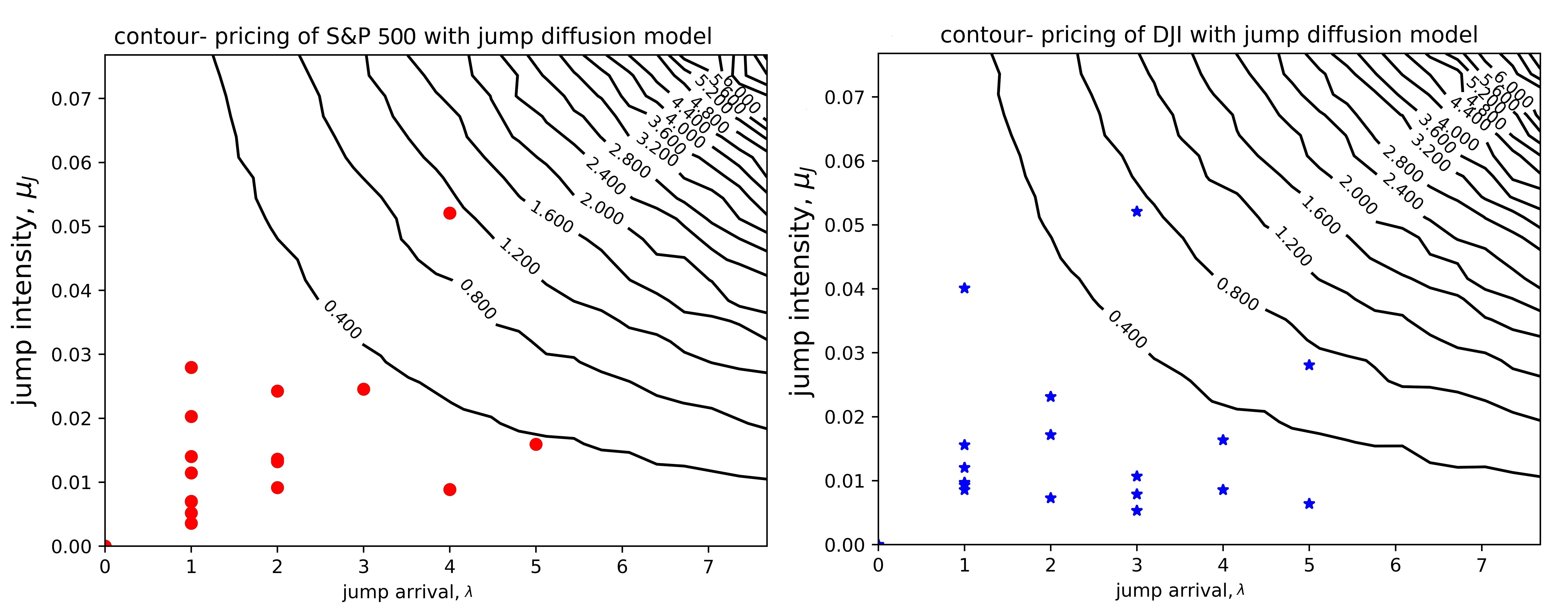}
\caption{The expected pricing for 20years of Dow Jones index and S\&P 500 index}\label{Fig6}
\end{figure}

On the right side of Figure \ref{Fig6}, the asterisk shows the expected positive jump arrivals and intensities for Dow Jones(DJI) over the pass 20years from 1995 to 2015. We could observe that, the expected payoff in pricing would differ from the European call,  where we did not consider jump at all. The highest attainable payoff could even reach approximately 0.7 for a normalized return.

Looking at the left side of Figure \ref{Fig6}, the dot shows the expected positive jump arrivals and intensities for S\&P 500 over the pass 20years from 1995 to 2015. We could observe that, the expected payoff in pricing not only differ from the European call but different as DJI indexes too. The highest attainable payoff could exceed 1.0 for a normalized return. This shows that, for different index or asset, the jump is inconsistent.

Next section we would try out the application of jump diffusion model in annuity pricing.
\newpage
\subsection{Annuity Pricing with Jump Diffusion Model}
As we mention in the Methodology, the pricing of annuity if indicate by Equation Eq.(\ref{Eqn8}). Eventhough pricing company are aware that the expected fair price of annuity are affected by the extreme events, they did not included the chances of risk in pricing.

\begin{figure}[ht]
\centering
\includegraphics[width=1.2\textwidth]{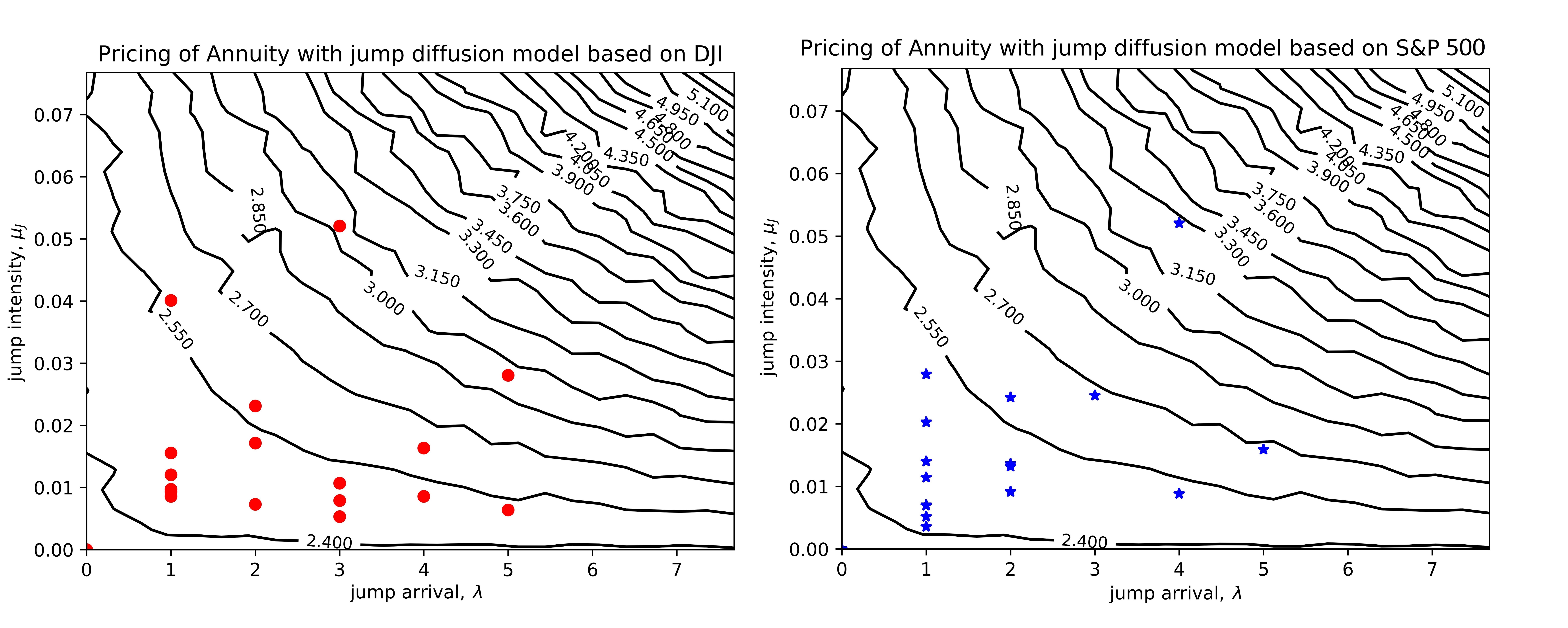}
\caption{The expected rewards function from the annuity with jump model based on DJI and S\&P 500.}\label{Fig7}
\end{figure}

Figure \ref{Fig7} shows that two different pricing payoff of annuity with jump diffusion model during positive upward jump event. The left figure shows the pricing payoff that is based on Dow Jones index, The red dots are historial yearly pricing from 1995 to 2015. On the right was based on S\&P 500 index, and same year period with DJI for the blue asterisks point.

When we observe closely at the left figure, the pricing of annuity without considering jump eventss are allocated at the origin with an payoff less than 2.4. However, we can see that most of the expected payoff from the historical data higher, when we expect a positive drifting jump event. This shows that without considering the chances of positive extreme events, the annuity is undervalued.

Vice versa, the expected payoff of pricing would be expecting a higher loss during a negatively drifted jump event. In results, causing the annuity policy to be overvalued. Either drift will resulting a fail in fair pricing. The need of consideration on jump event is important as the pricing company would lose its competitiveness to the other company with a over and undervalue price..

\newpage
\section{Conclusion and Discussion}
Both of the economic downturn in 1998, subprime crisis in year 2007 are the most noticeable extreme events happens in the economy. Undeniable that extreme event do exist in the market or even single asset. The ability to capture its signal before it happen is significant and important as it would definitely reduce the risk being bearded by investors. 
	
Jump diffusion model definitely more useful in capturing such signal than observing the past prices or market behavior such as moving average and trend. As for Black-Scholes model, we had shown that the inability to capture the extreme jump event would lead the investor into a riskier situation that BS model could not foresee.

This research had identify the effect of jump on European call options, on different jump intensity and occurrence. The results shows that the drift of jump, $\mu_J$ and arrival of jump, $\text{jump}_{\sigma}$ will affect the option payoff. The result will be either higher or lower depending on the drift of jump. Even the drift is zero, the payoff would be affected too, as the volatility of jump would alters the prices. This means that, whenever we expect a jump from the stock or market, even the drift is small, the risk of volatility is still exist. As long as there is a jump, there will be larger risk and a fluctuation of price should be expected.

Gibbs sampling technique could calibrate out the values and parameters of the market data, which is the drift and volatility of the model itself, the jump arrival and the drift and volatility of the jump. However, it had some difficulties, as the converging of parameters had some limitations. Sometimes the same path and pattern of a stock could be simulated using different set of parameters, which could be near to each other. For example, a high drift low volatility stock’s path might be able to attain by a slightly lower drift but higher volatility. Hence, the results attain had a range of possibilities. 

However,if the jump arrival and intensity is checked to be high, the stock itself almost surely is undergoing several jump event and a huge fluctuation in price is expected in the near future. The risk

	Jump diffusion model does not solely fits in options only. It could help in calculating the expected of other derivatives. The payoff of annuity is shown in results that, the reward function could be higher if the underlying asset are undergoing a positive jump event.  Similarly, a large risk of lost if it is going to undergoa negative jump event. In this case, we can say that the annuity is undervalued, as the risk being undertaking is far more larger than what they expect. For example, we refer back to Figure \ref{Fig7} from the previous section, the price should be somewhere around 2.6 rather than 2.3 which without considering the risk of jump.

	There could be other derivatives and purposes that jump diffusion model can fit into it for better risk managing and lost preventing. Hence, it is important to consider jump model than geometric Brownian motion when dealing with market derivatives.


\end{document}